\documentclass[aps,prb,twocolumn,groupedaddress,showpacs]{revtex4}
\usepackage{blkarray}
\usepackage{graphicx}% Include figure files
\usepackage{dcolumn}% Align table columns on decimal point
\usepackage{bm}% bold math
\usepackage{amsmath,amsfonts,amssymb,amsthm,verbatim}
\usepackage[ansinew]{inputenc}
\usepackage[usenames]{pstricks}
\usepackage{color}
\usepackage{epsfig}
\usepackage{MnSymbol}
\usepackage{amssymb}
\usepackage{mathrsfs}
\newcommand{\be}{\begin{equation}}
\newcommand{\ee}{\end{equation}}
\newcommand{\bea}{\begin{eqnarray}}
\newcommand{\eea}{\end{eqnarray}}

\newcommand{\bi}{\begin{itemize}}
\newcommand{\ei}{\end{itemize}}
\newcommand{\bc}{\begin{center}}
\newcommand{\ec}{\end{center}}

\begin{document}

\title{Transport through periodically driven systems:\\
Green's function approach formulated within frequency domain}

\author{Oleksandr Balabanov}
\affiliation{Department of Physics, University of Gothenburg, SE 412 96 Gothenburg, Sweden}

\begin{abstract}

The steady-state electronic transport across periodically driven systems can be efficiently addressed using Landauer-B\"{u}ttiker formalism. The time-dependent nonequilibrium Green's function theory then may be adapted for developing direct and universal calculation schemes. Here we propose an alternative scheme to carry out the calculations. The idea is based on treating the transport problem in frequency domain and designing Green's functions for the corresponding Hamiltonians rewritten in Floquet-Sambe formalism. We show that within our approach the expressions for time-periodic currents and densities essentially replicate the well known formulas from time-independent theory. The results are then simplified for easier implementation in numeric computations.
\end{abstract}

%\date{\today}
\maketitle

\section{Introduction} 

Over the decades there has been considerable progress in developing efficient methods for addressing steady-state stationary transport in mesoscopic and molecular devices. Numeric computations exploiting transport theories embedded within a nonequilibrium Green's function formalism are by now routine in many areas of quantum physics and chemistry. At present there exist several software packages \cite{soft1,soft2,soft3, soft4} offering environments for implementing calculations, with an enormous number of works having been conducted in this field.  

In contrast, numerical approaches for describing transport under time-dependent drives are far less developed, even though the basic foundation towards efficient algorithms for this task has already been laid \cite{Kurth_DFT, TR_method, TR_method2, TR_method3}. While transient and steady-state time-dependent transport across various systems have been successfully studied directly from first principles~\cite{dft_ex1, dft_ex2, dft_ex3, dft_ex4} or using tight-binding \text{theory}~\cite{kwant_ex1, kwant_ex2, kwant_ex3}, the problem is in general very hard to handle and typically requires massive computational power.

In this article we address long-time dynamics in periodically driven systems and propose a transparent method for describing time-periodic steady-state transport, with interactions treated self-consistently at a mean-field level. The idea is to formulate the problem within frequency domain, also called Floquet-Sambe space, and combine it with nonequilibrium Green's function formalism. Similar approaches have been taken in \text{Refs.~[16, 17]}, however, here we pattern the calculations on time-independent theory instead of exploiting time-dependent (Keldysh) formalism and by this facilitate a very easy passage to periodically driven transport. In fact, it is shown that the expressions for multiterminal current and charge density essentially replicate well known time-independent formulas. Importantly, we also focus on the computational aspects of the obtained expressions and exploit periodicity of the Floquet-Sambe space to bring them into numerically efficient form. It follows that by using our construction existing numerical self-consistent algorithms built for describing time-independent steady-state transport are %anticipated to be 
directly adaptable for handling also periodically driven systems. We provide a brief comparison between these algorithms and also discuss the feasibility of extending our computational scheme to go beyond the mean-field description by means of adapting concepts from time-dependent Density Functional Theory.

\begin{figure} \centering
    \includegraphics[width=8.5cm,angle=0]{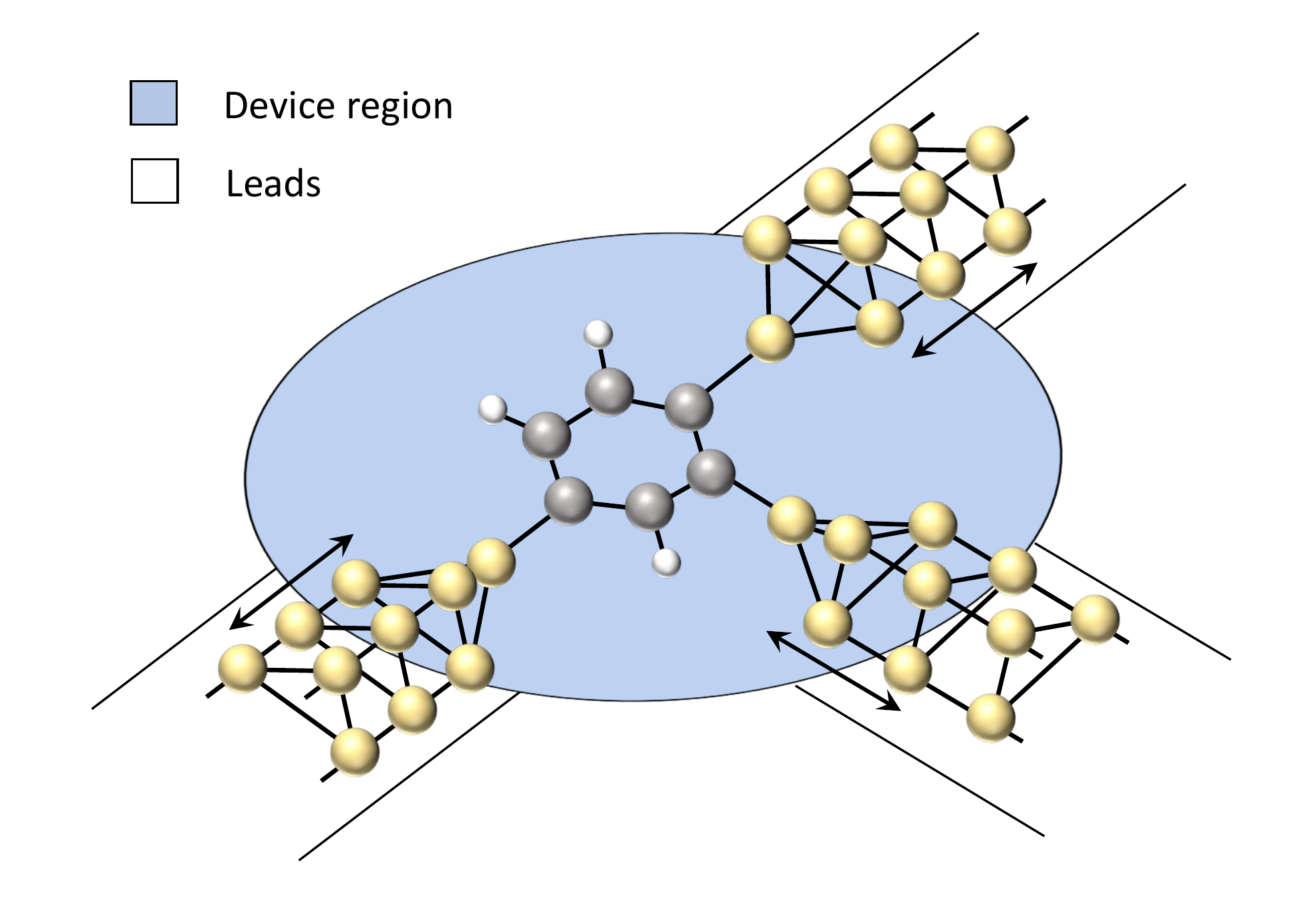}
    \caption{A schematic illustration of the setup: semi-infinite leads (white) connected to a device region (blue). The arrows represent couplings between the regions.}
     \label{fig1}
\end{figure}

The transport setup considered in this work consists of a finite scattering region (also called device, transport or central region) attached to an arbitrary number of semi-infinite leads, Fig.~1. We do not impose any restrictions on the form of the driving except that it is periodic in time. The driving may be implemented as, for example, a periodically varying external capacitive gate bias of the transport region, a dipole-field interaction with a laser field, ac applied voltages, or any combination between them with the only requirement that their frequencies are commensurable, i.e. all the drivings are periodic in time with some fundamental frequency $\Omega$.

The article is planned as follows. After a short introduction to the Floquet-Sambe formalism, we introduce the concept of a Green's function defined within frequency domain and describe its basic properties. In~Sec.~III we turn to transport theory formulated within this formalism and describe transport across the setup composed of a device region connected to leads. In the same section we discuss how one may incorporate interactions into the theory by solving the time-dependent Poisson equation, with possible extension to time-dependent Density Functional Theory. We compare the results with analogous ones from time-independent formalism in Sec.~IV. We conclude our findings in the summary section. The detailed derivations of all expressions for currents and densities can be found in the Appendices.

\section{Green's functions within Floquet-Sambe formalism}

\subsection{Floquet-Sambe Construction}

The periodically driven Schr\"{o}dinger equation can be formally addressed using the Floquet theorem. It \text{implies} the existence of a complete set of solutions of the form $\text{$|\psi(t) \rangle = e^{-i E t/\hbar} |u(t) \rangle$}$ where $E$ is a quasienergy defined $\text{modulo}$ $\hbar \Omega = 2\pi\hbar/T$ ($T$ is the driving period) and $|u(t) \rangle = |u(t + T) \rangle$ are time-periodic modes. By plugging in these modes to the Schr\"{o}dinger equation one arrives at the 
\begin{align}
\begin{split}
\left(H(t) - i \hbar \partial_t \right) |u(t)\rangle  = E |u(t)\rangle, 
\end{split}
\label{eq:Floquet_SE}
\end{align}
where both $H(t)$ and $|u(t)\rangle$ are time-periodic with the same period~$T$. One can reformulate Eq.~(\ref{eq:Floquet_SE}) in the extended Hilbert space $\mathscr{H} \otimes \mathscr{T}$ (sometimes called Floquet-Sambe space or frequency domain) consisting of all time-periodic states \cite{Sambe}, typically denoted with~$|\, ... \, \rrangle$. The equation one then gets is an eigenvalue problem within the extended Hilbert space, given by
\begin{align}
\begin{split}
\mathcal{H} |u\rrangle  = E |u \rrangle,
\end{split}
\label{eq:Fl}
\end{align}
where the infinite-dimensional matrix $\mathcal{H}$ represents the time-periodic Hamiltonian $H(t)$ within this formalism. Decomposed in a Fourier basis of all time-periodic states, i.e a basis of the form $|i, m \rrangle = e^{i  m \Omega t}|i\rangle$ where the orthogonal set $|i\rangle$ spans the conventional Hilbert space $\mathscr{H}$, it reads \cite{Shirley}
\begin{align}
\begin{split}
\mathcal{H} =  \begin{pmatrix}
 \ddots & \vdots & \vdots  & \vdots  & \udots\\
  ...  & H^{(0)} - \hbar\Omega \text{I} & H^{(1)} & H^{(2)} & ...  \\  
 ... & H^{(-1)} & H^{(0)} & H^{(1)} & ... \\ 
...  & H^{(-2)} & H^{(-1)} & H^{(0)} + \hbar\Omega \text{I} & ...  \\ 
  \udots  & \vdots  & \vdots  & \vdots  & \ddots\\
\end{pmatrix},
\end{split}
\label{eq:Sambe_SE2}
\end{align}
with \text{entries} $\llangle i, m| \mathcal{H} |j, m^\prime \rrangle =  \langle i| H^{(m^\prime - m)} + m \hbar \Omega I \delta_{m, m^\prime}   |j \rangle$ where $\langle i| H^{(n)} |j \rangle = 1/T \int^T_0 \, dt \, e^{i n \Omega t} \langle i| H(t) |j \rangle$ and $I$ is identity operator. Here we will refer to the indices $m$ and $m^\prime$ of any Floquet-Sambe matrix as \textit{Floquet-Sambe row and column indices} respectively. Note that $(m, \, m^\prime)$ divide the Floquet-Sambe matrix into an infinite number of \textit{Floquet-Sambe blocks}, each of which acts on a vector space of the same dimension as the conventional Hilbert space $\mathscr{H}$, see Eq.~(\ref{eq:Sambe_SE2}). The Floquet-Sambe blocks are not independent: For any integer $n$ the block $(m, \, m^\prime)$ is equal to the block $(m + n, \, m^\prime + n)$ minus the diagonal matrix $n \hbar \Omega I$ in case $m=m^\prime$, where $I$ denotes the identity matrix.

For not too strong drivings the Floquet-Sambe eigenstates typically vanish rapidly with the Floquet-Sambe index \cite{Rudner} and this allows for very efficient numerical calculations by simply truncating the infinite-dimensional matrix $\mathcal{H}$ at some sufficiently large $m$ and~$m^\prime$. For instance, in a previous work \cite{Balabanov} it was enough to consider just six values ($m, \, m^\prime  = -2, ..., 3$) to reach very good convergence of the result, however, in practice even smaller numbers may work and one should always optimize this parameter.

\subsection{Floquet-Sambe Green's Function: \\
\ \ \ \ \ Definition and Basic Properties}

The Floquet-Sambe equation~(\ref{eq:Fl}), with~${\cal H}$ defined in~(\ref{eq:Sambe_SE2}), has the same structure as the time-independent Schr\"{o}dinger equation but for a newly defined system. In fact, it is equivalent to the time-independent Schr\"{o}dinger equation for an extended system consisting of particles interacting with a quantized  \text{external} field\cite{Shirley}. This similarity with time-independent formalism suggests that by using the Floquet-Sambe formalism one should be able to straightforwardly generalize all the concepts of stationary transport theory to periodically driven systems. This is the main message of the present work.

In analogy to time-independent Green's function formalism, the retarded Green's function $\mathcal{G}(\omega)$ defined for a time-dependent Hamiltonian in $\mathscr{H} \otimes \mathscr{T}$ satisfies the equation
\begin{equation}
(\omega^+ I - \mathcal{H}) \mathcal{G}(\omega) = I,
\label{eq:Greeen_Def_FS1}
\end{equation}
where $I$ is an identity operator and $\mathcal{H}$ represents $H(t)$ rewritten within $\mathscr{H} \otimes \mathscr{T}$, cf. Eq.~(\ref{eq:Sambe_SE2}). Here $\omega^{+} = \omega+ i\eta$ with infinitesimal $\eta >0$. It follows that the Floquet-Sambe blocks of the Green's function $\mathcal{G}(\omega)$ satisfy the relation: $\mathcal{G}^{(m +n, \, m^\prime+n)}(\omega +n\hbar\Omega) = \mathcal{G}^{(m, \, m^\prime)}(\omega)$ with the upper indices denoting the Floquet Sambe row and column indices. This can be easily retrieved from the periodicity of the Floquet-Sambe Hamiltonian~$\mathcal{H}$, Eq.~(\ref{eq:Sambe_SE2}).

By design  $\mathcal{G}(\omega)$ contains all information on the periodic drive and in this form the time-dependence is very easy to handle. In the same way as the Floquet-Sambe Hamiltonian $\mathcal{H}$ represents the time-periodic Hamiltonian $H(t)$, the Floquet-Sambe Green's function $\mathcal{G}(\omega)$ is simply a representation of the time-periodic retarded Green's function $G(t, t^\prime)$ within the extended Hilbert space $\mathscr{H} \otimes \mathscr{T}$. The two Green's functions are connected via the relation~\cite{Kitagawa, Eissing}
\begin{align}
\begin{split}
\mathcal{G}(\omega) =  \begin{pmatrix}
 \ddots & \vdots & \vdots  & \vdots  & \udots\\
  ...  & G(0, \omega+\hbar\Omega) & G(1, \omega) & G(2, \omega-\hbar\Omega) & ...  \\  
 ... & G(-1, \omega +\hbar\Omega)& G(0, \omega) & G(1, \omega - \hbar\Omega) & ... \\ 
...  & G(-2, \omega + \hbar\Omega) & G(-1, \omega) & G(0, \omega - \hbar\Omega) & ...  \\ 
  \udots  & \vdots  & \vdots  & \vdots  & \ddots\\
\end{pmatrix},
\end{split}
\label{eq:Sambe_Keldysh_Green_functions}
\end{align}
where $G(n, \omega)$ is the Fourier-transformed retarded Green's function $G(t,t^\prime)$ of time-dependent nonequilibrium formalism. For a quick consistency check we note that Eq.~(5) correctly reproduces the time-independent limit when we set the driving to zero; the off-diagonal terms just vanish in this case. In principle, any time-periodic operator can be recast in Floquet-Sambe form, one just has to Fourier transform the operator and combine the obtained components into a matrix of the same structure as in Eq.~(\ref{eq:Sambe_Keldysh_Green_functions}). This for example may be done for lesser, $G^<(t,t^\prime)$, and greater, $G^>(t,t^\prime)$, Green's functions~\cite{Eissing, Cao1},  however, in this case the construction is introduced in a rather artificial way and looses its intuitive meaning. %Even though we believe that one should be able to make use of Floquet-Sambe construction for these Green's functions, 
Here we choose to focus on the retarded Green's function $\mathcal{G}(\omega)$ only, defined in Eq.~(\ref{eq:Greeen_Def_FS1}), and derive all the quantities in terms of it. We show that the procedure essentially resembles familiar time-independent derivations and is easily adaptable to well-established approaches developed for describing transport in stationary systems.

A key element of any Green's function formalism is the Dyson equation allowing for effective handling of system partitioning. Directly from the definition, the Green's function $\mathcal{G}(\omega)$ satisfies the Floquet-Dyson equation 
\begin{align}
\begin{split}
\mathcal{G}(\omega) & = \mathcal{G}_0(\omega) + \mathcal{G}_0(\omega) \, \mathcal{V} \, \mathcal{G}(\omega),
\end{split}
\label{eq:Dyson_equation_SF}
\end{align}
with arbitrary partitioning of the time-periodic Hamiltonian $\mathcal{H} = \mathcal{H}_0 + \mathcal{V}$ and $\mathcal{G}_0(\omega) = [\omega^+ I - \mathcal{H}_0]^{-1}$. The Floquet-Dyson equation can also be derived using time-dependent Green's functions $G(t, t^\prime)$ \cite{Tsuji1, Joura, Tsuji2}, however, in the form of~Eq.~(\ref{eq:Dyson_equation_SF}), it reflects well the similarity between time-independent and periodically driven  theories, to be highlighted here.

A formal solution to Eq.~(\ref{eq:Greeen_Def_FS1}) can be expressed in terms of a full set of eigenstates $|k\rrangle$ with eigenvalues $E_k$, i.e. a set of states satisfying $\mathcal{H} |k\rrangle  = E_k |k \rrangle$. One finds
\begin{equation}
\mathcal{G}(\omega) = \sum_{k} \frac{|k\rrangle\llangle k|}{\omega - E_k + i \eta}.
\label{eq:Greeen_Sol_FS}
\end{equation}

The eigenstates $|k\rrangle$ exhibit a high degree of redundancy:  the states $|k\rrangle$ and $|k^\prime\rrangle$ with $E_k - E_{k^\prime} = n \hbar \Omega$ ($n \in \mathbb{Z}$) represent the same physical state and can be expressed in terms of each other. This affects the Green's function $\mathcal{G}(\omega)$ that contains lots of repetitive information, cf. Eq.~(\ref{eq:Sambe_Keldysh_Green_functions}). The same argument also applies to the Floquet-Sambe spectral function~$\mathcal{A}(\omega)$, \text{defined~as}
\begin{align}
\begin{split}
\mathcal{A}(\omega) = i[\mathcal{G}(\omega) - \mathcal{G}^\dagger(\omega)] = 2 \pi \sum_k \delta(\omega - E_k) |k\rrangle \llangle k|. 
\end{split}
\label{eq:Spectral_Def_FS}
\end{align}

As discussed later in this work, a careful treatment of such redundancy is critical for avoiding overcounting in calculations of various quantities, most notably currents and densities.

\section{Transport Across Periodically Driven System}

\subsection{Time-independent Transport: Review}

 To provide a background, we first discuss results from time-independent transport theory. Although the theory is well known, we here go into some detail so as to facilitate the passage to Floquet-Sambe time-dependent transport. Consider a setup commonly assumed within Landauer-B\"{u}ttiker formalism: a~finite device region connected to a number of semi-infinite leads (labeled by an index~$l$) biased by applied voltages $V_l$, see Fig.~1. \text{Usually},  several surface atomic layers of the physical leads at the interface to the device are included in the device region so that the leads are treated as if composed entirely of unperturbed space-periodic bulk. The equilibrium and biased leads are assumed to be described by~Hamiltonians $H^\text{eq}_l$ and  $H_l = H^\text{eq}_l + e V_l$ respectively, the \text{(extended)} device region by~$H_d$, and the tunneling between leads and device by~$H_{l, d}$. It is also considered that there is no direct coupling between the leads, and that the spin degree of freedom is included in the Hamiltonian.

The \text{retarded} Green's function $G(\omega)$ fulfills the relation
\begin{equation}
 (\omega^{+} I - H) G(\omega) = I,
 \label{eq:Greeen_Def}
\end{equation}
where $H$ is the full Hamiltonian, $H= H_l + H_d + H_{l,d}$, $I$ is the identity, and $\omega^{+} = \omega + i\eta$ with infinitesimal positive $\eta$. Using simple block by block matrix multiplications one easily derives the following formula for the device region's Green's function:
\begin{align}
\begin{split}
    G_{d}(\omega) = \left[ \omega^+ I_d - H_d - \sum_l \Sigma_{l} \right]^{-1},\\
\end{split}
\label{eq:block_eq3}
\end{align}
with self-energies $\Sigma_{l} = H_{d,l} \, G_l \, H_{l, d}$. Here $G_l$ is the $\text{$l$-th}$ lead Green's function, i.e. $G_l(\omega) = [\omega^+ I_l - H_l]^{-1} = G^\text{eq}_l(\omega - eV_l)$ where $G^\text{eq}_l(\omega)$ is the surface Green's function of the unbiased lead. The self-energies $\Sigma_{l}$ can be obtained from the surface Green's function of the semi-infinite leads, i.e. the part of $G_l(\omega)$ corresponding to sites close to the surface. In practice, the surface Green's function can be found iteratively or by exploiting eigenstate decomposition. In its turn, the device region's Green's function $G_{d}(\omega)$ can be computed by direct inversion of Eq.~(\ref{eq:block_eq3}), or, more efficiently, by using recursive Green's function techniques \cite{recursive1, recursive2, recursive3}. 

The retarded Green's function $G_{d}(\omega)$ may be used for computing key observables including steady-state currents and electron densities. The dc current $I^{\text{dc}}_l$ in lead $l$ can be calculated exploiting the Landauer-B\"{u}ttiker formula with lead to lead transmissions expressed within a nonequilibrium Green's function formalism~\cite{Datta}. There are two common approaches for deriving this relation. One of them expresses the current in terms of lesser Green's function and then transforms the expression using the Langreth rules~\cite{Meir}. Alternatively, one employs the Lippmann-Schwinger equation for explicit calculation of the Landauer-B\"{u}ttiker scattering states~\cite{soft1, Todorov, Paulsson}. This latter approach just requires knowledge of the retarded Green's function and, as we shall see in the next subsection, is easily generalizable to time-periodic systems through the Floquet-Sambe approach.

The procedure is straightforward: We express scattering states in terms of $G_{d}(\omega)$ and then populate them according to the Landauer-B\"{u}ttiker formalism, i.e. the scattering states are considered to be in equilibrium with the corresponding lead. It is assumed that each lead is kept at a chemical potential $\mu_l$, temperature $T_l$, and applied dc voltage $V_l$. The technical details are collected in Appendix~A. The final result is the already mentioned Landauer-B\"{u}ttiker formula~\cite{Datta}
\begin{align}
\begin{split}
I^\text{dc}_{l} =  \frac{e}{h} \sum_{l^\prime \neq l} \int_{-\infty}^{\infty} \, d\omega\,  &T_{l^\prime, l} (\omega) \, \big ( f_{l} (\omega  - eV_{l}) -  f_{l^\prime} (\omega -  eV_{l^\prime}) \big ),
\end{split}
\label{eq:current_total_TI}
\end{align}
with transmissions given by $T_{l, l^\prime} = \text{Tr}  \left[ G^\dagger_d \Gamma_{l^\prime} G_d \Gamma_l \right]$. %known as the Fisher-Lee relation~\cite{Fisher}. 
Here $f_l(\omega)$ is a Fermi-Dirac distribution function with chemical potential $\mu_l$ of the \text{$l$-th} lead and $\Gamma_{l} = i(\Sigma_{l} - \Sigma_{l}^\dagger)$ is a so-called broadening function~\cite{Datta}. This expression is computationally efficient: The integral can be evaluated by integrating over only a narrow energy window where the difference between the Fermi functions is not negligible.

A similar path can be followed for calculating the electron density $\rho(x)$, important for including interactions self-consistently. The density matrix of the device region is expressed as follows~\cite{soft1, soft2, Paulsson} (see Appendix~A):
\begin{align}
\begin{split}
D_d = \frac{1}{2\pi} \sum_l \int^\infty_{-\infty} d\omega f_l(\omega - eV_l) [G_d  \Gamma_l G^\dagger_d ](\omega).
\end{split}
\label{eq:Density_FS}
\end{align}
The device region's density $\rho_d(x)$ is then obtained by evaluating the density matrix at equal positions, $\rho_d(x) = D_d(x,x)$. Without loss of generality, here we switched to the position basis~$|x\rangle$ instead of the general basis~$|i\rangle$ assumed previously.

The expressions for current and density matrix, Eqs.~(\ref{eq:current_total_TI}) and (\ref{eq:Density_FS}), take into account only scattering states, i.e. current carrying modes originating in one of the leads, and completely ignore localized (bound) states. These latter states do not directly contribute to the measured current, however, may affect it through interactions and therefore have to be included into the formalism. The occupation of bound states depends crucially on the system's dynamics before reaching the steady state. Generally speaking, this implies that the occupation function of bound states has to be externally specified from the experiment. However, these states are typically outside of the narrow conduction window, $E_\text{bound} < \text{min}(\mu_l + eV_l)$ or $ E_\text{bound} > \text{max}(\mu_l + eV_l)$, and if we assume that the applied voltages were switched on slowly enough, then the bound states are anticipated to be filled up according to some equilibrium Fermi function $f_\text{eq}(\omega)$ with equilibrium chemical potential $\mu_\text{eq}$. Bound states may then be included into the formalism through realizing the procedure outlined in Refs.~[1, 2]. The idea here is to separate the density matrix into two parts, equilibrium $D^{\text{eq}}$ and nonequilibrium~$D^{\text{neq}}$. The equilibrium part takes the following form:
\begin{align}
\begin{split}
D_d^{\text{eq}} &= \int_{-\infty}^{\infty} \, d\omega \, f_\text{eq}(\omega) \sum_k \delta(\omega-E_k)
| u_{d,E_k} \rangle \langle u_{d,E_k} |\\
& = -\frac{i}{2\pi}\int_{-\infty}^{\infty} \, d\omega \, f_\text{eq}(\omega) [G_d^\dagger(\omega) - G_d(\omega)].
\end{split}
\label{eq:Density_FS_eq}
\end{align}
This expression includes scattering as well as localized states and populates them according to $f_\text{eq}(\omega)$. The nonequilibrium part is calculated by simply subtracting Eq.~(\ref{eq:Density_FS_eq}) from Eq.~(\ref{eq:Density_FS}): 
\begin{align}
\begin{split}
D_d^{\text{neq}} &= \frac{1}{2\pi} \sum_l \int^\infty_{-\infty} d\omega \{f_l(\omega - eV_l)-f_\text{eq}(\omega)\} [G_d  \Gamma_l G^\dagger_d ](\omega)
\end{split}.
\label{eq:Density_FS_eq2}
\end{align}

In general, the integral in Eq.~(\ref{eq:Density_FS}) is difficult to compute numerically because it has to be evaluated on a very fine grid (since the poles are very close to the real axis) and, moreover, it is unbounded. On the other hand, the computation of the contributions $D_d^{\text{eq}}$ and~$D_d^{\text{neq}}$ is much simpler: the equilibrium part can be calculated by a contour integration and the nonequilibrium part is bounded by two Fermi functions. A detailed discussion of these computational approaches may be found in Refs.~[1, 2]. Thus, even in the absence of bound states, the separation of the total density matrix into equilibrium and nonequilibrium components is vitally important. In the next subsection we will show - by using a Floquet-Sambe approach - that a procedure similar to that for time-independent transport, reviewed above, can be constructed for time-periodic steady-state transport. This new approach has the potential to significantly reduce the cost of numerical computations.

\subsection{Time-dependent Transport: Currents and Densities}

Following in the footsteps of the time-independent Green's function formalism, we will now develop analogous concepts and results within Floquet-Sambe space. We take the same setup as in the previous section, see Fig.~1, but now allow the terms to be generically periodic in time: the leads are described by~$H_l(t)$, the (extended) device region by~$H_d(t)$, and tunnelings by~$H_{l, d}(t)$. The system is partitioned in such a way that there are no direct coupling elements between the leads.

The time-dependence of the leads is assumed to come only from time-periodic applied voltages~$V_l(t)$. The typical leads in experiments are good metals and $V_l(t)$ may therefore be assumed to alter only the single-particle energies of incoming states and to have no effect on statistics~\cite{Pedersen_TFS, Moskalets_TFS1}. It follows that the time-dependent part $V^\text{TD}_l(t)$ of the applied voltage~$V_l(t) = V_{l}^\text{DC} + V^\text{TD}_l(t)$ can be gauged out at the cost of introducing extra time-periodic phases to the couplings~$H_{l, d}(t)$. The transformed couplings are explicitly given by~\cite{Arrachea_TFS1, Shevtsov}
\begin{align}
\begin{split}  
H^\text{new}_{l, d}(t) = \exp\left[\frac{i}{\hbar} \int^t_0 \, dt^\prime \, e \,  V^\text{TD}_l(t^\prime)\right]H_{l, d}(t).
\end{split}
\label{eq:Density_FS_phase}
\end{align}
Note that the phases remain periodic in time after the transformation as long as the time-periodic $V^\text{TD}_l(t)$ does not include any dc voltage. By definition, any gauge transformation does not affect any observables and therefore, without loss of generality, from now on we may assume time-independent leads~$H_l(t) = H_l = H^\text{eq}_l + e V^\text{DC}_l$. %~\cite{Arrachea_TFS1}. 
The full time-dependent Hamiltonian $H(t)$ is then converted to Floquet-Sambe form $\mathcal{H}$, cf. Eq.~(\ref{eq:Sambe_SE2}). 

By exploiting the ideas presented in Sec.~II, we construct the Green's function $\mathcal{G}(\omega)$ within Floquet-Sambe space and then employ it for calculating steady-state transport properties. Similarly to the time-independent case, the Green's function in the device region takes the form
\begin{align} \begin{split} \mathcal{G}_{d} = \left[\omega^+ I_d - \mathcal{H}_d - \sum_l \mathcal{E}_l \right]^{-1},\\ \end{split} \label{eq:block_eq3_FS} \end{align}
where, in obvious notation, self-energies are defined through $\mathcal{E}_l = \mathcal{H}_{d, l} \mathcal{G}_l \mathcal{H}_{l, d}$. Here $\mathcal{G}_l (\omega)= [\omega^+ I_l - \mathcal{H}_l]^{-1}$ is the $l$-th lead Green's function given within Floquet-Sambe formalism. Since all leads are assumed to be stationary, recall that the time-dependent part of the applied voltages has been gauged out using Eq.~(\ref{eq:Density_FS_phase}), the $l$-th lead Green's function $\mathcal{G}_l$ takes the form of a block-diagonal matrix 
\begin{align}
\begin{split}
\mathcal{G}_l =  \begin{pmatrix}
 \ddots & \vdots & \vdots  & \vdots  & \udots\\
  ...  & G_l(\omega+\hbar\Omega) & 0 & 0 & ...  \\  
 ... & 0& G_l(\omega) & 0 & ... \\ 
...  & 0 & 0 & G_l(\omega - \hbar\Omega) & ...  \\ 
  \udots  & \vdots  & \vdots  & \vdots  & \ddots\\
\end{pmatrix},
\end{split}
\label{eq:Sambe_Keldysh_Green_functions_lead}
\end{align}
with blocks $\mathcal{G}^{(m, \, m^\prime)}_{l}(\omega) = G_l (\omega - m \hbar \Omega)\,\delta_{m, m^\prime}$ where  $G_l(\omega) = [\omega^+ I - H_l]^{-1} = G^\text{eq}_l(\omega - e V^\text{DC}_l)$ is the conventional Green's function of time-independent lead $l$ biased at~$V^\text{DC}_l$. Thus, $\mathcal{G}_l (\omega)$ can be straightforwardly constructed once $G_l(\omega)$ is known. Clearly, written in the form of Eq.~(\ref{eq:block_eq3_FS}), the Green's function~$\mathcal{G}_{d}$ just replicates the analogous equation for the time-independent Green's function $G_d$, implying that all methods developed for computing it can also be applied here, including the widely used recursive Green's function techniques~\cite{recursive1, recursive2, recursive3}.

The steady-state currents flowing across the system can be addressed analogously to the stationary case: We calculate the scattering states with the help of the Lippmann-Schwinger equation and then populate them accordingly. In general, the steady-state currents in periodically driven systems are also periodic in time and therefore for complete analysis all Fourier components of the time-periodic currents have to be calculated. In fact, each of them can be expressed using the Floquet-Sambe construction. Here, to make the results more transparent and easier to follow, we present only expressions for the dc (rectified) current component and put the discussion on other components in Appendix~B.  As shown in the same appendix, the dc component $I_l^{(0)}$ of the total current passing through lead $l$ can be expressed in the following way:
\begin{align}
\begin{split}
I_l^{(0)}  = \frac{e}{h} \sum_{l^\prime \neq l} \int_{-\infty}^{\infty} \, d\omega \, &  \,  
\bigg ( \mathcal{T}^{\, (0)}_{l, l^\prime} (\omega) f_{l}(\omega - eV_{l}) \\
&-  \mathcal{T}^{\, (0)}_{l^\prime, l} (\omega) f_{l^\prime}(\omega - eV_{l^\prime}) \bigg),\\ 
\end{split}
\label{eq:current_total_FS_dc}
\end{align}
where $\mathcal{T}^{\, (0)}_{l, l^\prime} (\omega) = \text{Tr} \left[\mathcal{G}^\dagger_d \varGamma_{l^\prime}  \mathcal{G}^{\phantom\dagger}_d  \varGamma^{(0)}_{l} \right]$ and $\varGamma_{l^\prime} = \mathcal{H}_{d, l^\prime}\mathcal{A}_{l^\prime}\mathcal{H}_{l^\prime, d}$, with $\mathcal{A}_{l^\prime} = i(\mathcal{G}_{l^\prime}- \mathcal{G}^\dagger_{l^\prime})$ the Floquet-Sambe analogue of the broadening function. Here we have also defined a new operator $\varGamma^{(0)}_{l} = \mathcal{H}_{d, l}  [\mathcal{K}^{0} \, \mathcal{A}_{l} ] \mathcal{H}_{l, d}$ with entries $\llangle j, m^\prime |\mathcal{K}^{0}| i, m\rrangle  = \delta_{i,j} \delta_{m, \, 0} \, \delta_{m^\prime, \, 0}$. The newly introduced operator $\mathcal{K}^{0}$ can be interpreted as follows: $\mathcal{K}^{0}$ is the Floquet-Sambe zero matrix with its $(m, \, m^\prime) = (0, 0)$ block being replaced by the time-independent identity matrix. This operator removes the state overcounting present in the Floquet-Sambe spectral function~$\mathcal{A}_{l}$, cf. Sec.~II. We note that our expression is in good agreement with the one derived using a Floquet-Keldysh formalism~\cite{Arrachea_TFS1} but takes a more compact form and also generalizes it to more general time-periodic transport systems.

The obtained result, Eq.~(\ref{eq:current_total_FS_dc}), closely resembles the analogous relation from the stationary case, cf. Eq.~(\ref{eq:current_total_TI}), and highlights the similarity between time-independent and Floquet-Sambe formalisms. There is one substantial difference though: In Eq.~(\ref{eq:current_total_FS_dc}) the Fermi functions are not subtracted from each other and therefore in general the integral has to be evaluated over an infinite (or very large) $\omega$-interval. In Appendix C we derive an alternative but equivalent expression for the current $I^{(0)}_l$ where this problem is avoided:
\begin{align}
\begin{split}
I_l^{(0)} =\ \frac{e}{h} \sum_{l^\prime}  \int_{-\infty}^{\infty} \,  d\omega  \,  \bigg( \, 
\mathcal{T}^{\, (0)}_{l, l^\prime} (\omega) f_{l}(\omega - eV_{l}) \, - \, &\mathcal{T}^{\, (0) \, F}_{l, l^\prime} (\omega)  \bigg),
\end{split}
\label{eq:current_total_FS_TRS_final_short}
\end{align}
with $\mathcal{T}^{ (0)}_{l, l^\prime} = \text{Tr} \left[\mathcal{G}^{\dagger}_d \varGamma_{l^\prime}  \mathcal{G}^{\phantom\dagger}_d  \varGamma^{(0)}_{l} \right]$ and $\mathcal{T}^{ (0) F}_{l, l^\prime} = \left[\mathcal{G}^{\phantom\dagger}_d \varGamma^F_{l^\prime}  \mathcal{G}^{\dagger}_d  \varGamma^{(0)}_{l} \right]$. The summation is over all leads $l^\prime$ including $l^\prime = l$. Here the operator $\varGamma^F_l (\omega)$ is an operator very similar to the Floquet-Sambe broadening function $\varGamma_l (\omega)$ but with one important difference: $\varGamma^F_l (\omega)= \mathcal{H}_{d,l} [\mathcal{F}_l \mathcal{A}_l] \mathcal{H}_{l,d}$, where $\llangle i, m|\mathcal{F}_l(\omega)|j, m^\prime \rrangle = f_l (\omega - eV_l - m \hbar\Omega ) \delta_{i, j}\delta_{m, m^\prime}$ \text{is} a representation of the Fermi-Dirac distribution function $f_l (\omega - eV_l)$ within Floquet-Sambe space \cite{Cao1, Eissing}. The matrix $\mathcal{F}_l(\omega)$ can be viewed as a block-diagonal matrix with each Floquet-Sambe diagonal block $(m,\, m)$ constructed by multiplying the Fermi function $f_l (\omega - eV_l - m \hbar\Omega )$ with an identity matrix. The advantage of introducing $\varGamma^F_{l}$ is that the truncation of the Floquet-Sambe space comes with an added bonus when performing the integration. While the infinite-dimensional Fermi-Dirac matrix $\mathcal{F}_l(\omega)$ is never constant or vanishes with $\omega$, once truncated it reduces to identity (zero matrix) at sufficiently small (large) energies~$\omega$, i.e. at energies where \text{$f_l (\omega - eV_l + m_\text{max} \hbar\Omega )\simeq1$}  or $f_l (\omega - eV_l - m_\text{max} \hbar\Omega )\simeq 0$ with the cutoff~$m_\text{max}$. In Appendix C we show that in these two cases 
\begin{align}
\begin{split}
\sum_{l^\prime}\mathcal{T}^{ (0)}_{l, {l^\prime}} f_{l}(\omega - eV_{l}) \simeq \sum_{l^\prime} \mathcal{T}^{ (0) F}_{l, l^\prime}
\end{split}
\label{eq:current_total_FS_TRS_final_short2}
\end{align}
and therefore the integral in Eq.~(\ref{eq:current_total_FS_TRS_final_short}) has to be computed just over the region where the truncated $\mathcal{F}_l(\omega)$ differs from zero or identity, i.e. evaluated over a finite interval bounded by the applied voltages $V_l$ and Floquet-Sambe cutoff $m_\text{max}$. This significantly simplifies the computations.

In exact analogy to the time-independent case we may also calculate the time-periodic density matrix $D(t)$ using the Floquet-Sambe construction. The device region's density matrix rewritten within Floquet-Sambe space, call it $\mathcal{D}_d$, takes the form \text{(see Appendix~B)} 
\begin{align}
\begin{split}
\mathcal{D}_d =  \frac{1}{2\pi}\sum_l \int^\infty_{-\infty} d\omega [\mathcal{G}_d  \varGamma^F_l \mathcal{G}^\dagger_d] (\omega),
\end{split}
\label{eq:Density_FS_Sambe}
\end{align}
where $\varGamma^F_l (\omega) = \mathcal{H}_{d,l} [\mathcal{F}_l \mathcal{A}_l] \mathcal{H}_{l,d}$ has been defined earlier. It is here instructive to do a quick consistency check: In the time-independent limit all matrices are diagonal in the Floquet row and column indices and each Floquet diagonal block correctly reduces to the time-independent density matrix given in Eq.~(\ref{eq:Density_FS}). The Floquet-Sambe matrix elements of the Fourier transformed time-periodic density $\rho_d(x,t)$, call them $\varrho^{(m, \, m^\prime)}_d(x)$, are then given by $\varrho^{(m, \, m^\prime)}_d(x) = \llangle x, m |\mathcal{D}_d | x, m^\prime \rrangle$ where $|x, m \rrangle = e^{i m \Omega t}| x\rangle$ represents the position basis within Floquet-Sambe space.

We now outline a calculation scheme for the density matrix $\mathcal{D}_d$ that is more computationally efficient than a direct calculation through Eq.~(\ref{eq:Density_FS_Sambe}). In analogy to the time-independent case, we aim to split the total density matrix into two computationally easier parts. Within Floquet-Sambe space the first part is defined as
\begin{align}
\begin{split}
\mathcal{D}_d^\text{I} &=  -\frac{i}{2\pi}\int_{-\infty}^{\infty} \, d\omega \, f_\text{0}(\omega) [\mathcal{G}_d^\dagger(\omega) - \mathcal{G}_d(\omega)] \\
&= \int_{-\infty}^{\infty} \, d\omega \, f_\text{0}(\omega) \sum_k \delta(\omega-E_k)
| u_{d,E_k} \rrangle \llangle u_{d,E_k} |, \\
\end{split}
\label{eq:Density_FS_eq_trick}
\end{align}
with a Fermi function $f_\text{0}(\omega)$ corresponding to some chemical potential $\mu_0$. Even trough the matrix  $\mathcal{D}_d^\text{I}$ is defined within Floquet-Sambe space, it does not represent a time-periodic matrix in conventional Hilbert space since it does not possess the structure of a Floquet-Sambe matrix, see Sec.~II. Nevertheless, it populates $| u_{d,E_k} \rrangle$ according to the Fermi function $f_\text{0}(\omega)$, but now the steady states with different quasienergies and corresponding to the same physical state are populated differently. Note that both scattering and bound steady states are present in the summation of Eq.~(\ref{eq:Density_FS_eq_trick}) and therefore included in~$\mathcal{D}_d^\text{I}$.

To obtain the second contribution to the total density matrix, denote it by $\mathcal{D}^\text{II}_d$, the scattering states are to be subtracted from Eq. (\ref{eq:Density_FS_Sambe}). It follows that
\begin{align}
\begin{split}
\mathcal{D}^\text{II}_d = \frac{1}{2\pi}\sum_l \int^\infty_{-\infty} d\omega [\mathcal{G}_d  \{\varGamma^F_l - f_\text{0} \varGamma_l \} \mathcal{G}^\dagger_d] (\omega). 
\end{split}
\label{eq:Density_FS_eq_trick2}
\end{align}

If there are no Floquet bound states present in the system, the contributions $\mathcal{D}_d^\text{I}$ and $\mathcal{D}_d^\text{II}$ together yield an exact representation of the density matrix. Different from Eq.~(\ref{eq:Density_FS_Sambe}), they are expected to be much easier to compute numerically: The truncated $\mathcal{D}_d^\text{I}$ can be computed by the same contour integration as the equilibrium density $D_d^\text{eq}$ in the time-independent theory~\cite{soft1,soft2}. As a further advantage, the integral representing $\mathcal{D}^\text{II}_d$ is bounded by two Fermi functions, $f_\text{0}(\omega)$ and the truncated $\mathcal{F}_l(\omega)$. Thus, such a decomposition is expected to be very useful for numerical implementation. Note that the truncation spoils the Floquet-Sambe periodicity of the density matrix $\mathcal{D}_d$ in the Floquet index and therefore has to be restored by hand after each iteration step.

Similar to the time-independent case, the expression for density matrix, Eq.~(\ref{eq:Density_FS_Sambe}), does not take into account Floquet bound states, i.e. localized time-periodic eigenmodes of Floquet-Sambe Hamiltonian $\mathcal{H}$. To the best of our knowledge, there has so far been no systematic study describing the effect of Floquet bound states on transient or long-time transport across periodically driven systems. We shall leave the details of this problem open, and here only discuss it in very general terms. In general, to find the steady-state occupations one has to solve a set of rate equations describing energy exchange with a weakly coupled thermal bath~\cite{bound_states_F_Kohn1, bound_states_F_topological}. Within a time-independent formalism the resulting state occupations are described by a Boltzmann distribution independently of the form of system-bath coupling~\cite{desnsity_book}.  This is not so for periodically driven open systems. There is no generic steady-state distribution describing a Floquet steady-state and it has to be determined on a case-by-case basis. Nevertheless, for weak driving the steady-state distribution is expected to be close to the time-independent distribution~\cite{bound_states_F_Kohn2} where all energies which appear in the Boltzmann weights are replaced by corresponding quasienergies closest to the average energies. Thus, the bound states with average energy much below (above) the chemical potential can be assumed to be occupied (empty) with a good accuracy for not too strong driving fields. 

Now recall that the bound states are accounted for in~$\mathcal{D}_d^\text{I}$ and, in general, violate the quasienergy translational invariance there. However, the truncation of the infinite-dimensional matrix $\mathcal{F}_l(\omega)$ has an important consequence: Well below the chemical potential, in the region where $f_\text{0}(\omega + m_\text{max} \hbar \Omega)$ is approximately unity (here $m_\text{max}$ denotes a numerical cutoff), the steady states $| u_{d,E_k} \rrangle$ are all populated and therefore fulfill the quasienergy translational invariance. Thus, such modes are properly accounted for. Clearly, average energies of these modes also lie much below the chemical potential. This implies that the bound states well below (above) the chemical potential, i.e. in the region where $f_\text{0}(\omega + m_\text{max} \hbar\Omega)$ is close to unity ($f_\text{0}(\omega - m_\text{max} \hbar \Omega)$ is vanishing), are populated (empty) and correctly reproduce the limiting case discussed above. If the bound states are populated in a different way or some of them have average energies close to the chemical potential a different approach must be taken. 

\subsection{Time-dependent Transport: Self-consistent treatment of Coulomb Interaction}

The possibility to include Coulomb interaction is vital for any theory intended to capture time-periodic transport properties. Periodically driven systems treated within a non-interacting formalism violate two fundamental laws of electronic transport: {\text{conservation}} of current and gauge invariance of the applied voltages~\cite{Buttiker_AC, Buttiker_AC2}. These issues are connected to the presence of displacement currents arising across the system due to the time-dependent nature of the drive. On the other hand, taking the electron-electron interaction into account within a self-consistent theory resolves both problems \cite{Buttiker_AC, Buttiker_AC2, Buttiker_AC3, Wang} and here we discuss how one may design such a self-consistent procedure by exploiting the Floquet-Sambe construction.

A minimal self-consistent treatment of electron-electron interactions implies solving a time-dependent Poisson equation and then modifying a single-particle Schr\"odinger equation according to the obtained Coulomb potential $U(x,t)$. The time-dependent Poisson equation reads as follows:
\begin{align}
\begin{split}
\nabla \cdot \left(\epsilon(x) \nabla  U(x,t)\right) =  -e \rho(x, t),
\end{split}
\label{eq:Poison_Eq}
\end{align}
where $\epsilon(x)$ is in general a space-dependent dielectric constant and $e \rho(x, t)$ is the electron charge density. The boundary condition is taken such that $U(x, t)$ equals the applied voltage $V_l (t)$ sufficiently deep in the $l$-th lead. The Poisson equation can be Fourier transformed with respect to time and rewritten in Floquet-Sambe space as
\begin{align}
\begin{split}
\nabla \cdot \left(\epsilon(x) \nabla \, \mathcal{U}(x) \right) = - e \varrho (x),
\end{split}
\label{eq:Poison_Eq_FS}
\end{align}
with $\mathcal{U}(x)$ and $\varrho (x)$ taking the place of $U(x, t)$ and $\rho(x, t)$ within Floquet-Sambe theory. Explicitly:
\begin{align}
\begin{split}
&\mathcal{\varrho} =  \begin{pmatrix}
 \ddots & \vdots & \vdots  & \vdots  & \udots\\
  ...  & \rho^{(0)} & \rho^{(1)} & \rho^{(2)} & ...  \\  
 ... & \rho^{(-1)}& \rho^{(0)} & \rho^{(1)} & ... \\ 
...  & \rho^{(-2)} & \rho^{(-1)} & \rho^{(0)} & ...  \\ 
  \udots  & \vdots  & \vdots  & \vdots  & \ddots\\
\end{pmatrix}
\text{,} \\
&\mathcal{\mathcal{U}} =  \begin{pmatrix}
 \ddots & \vdots & \vdots  & \vdots  & \udots\\
  ...  & U^{(0)} & U^{(1)} & U^{(2)} & ...  \\  
 ... & U^{(-1)}& U^{(0)} & U^{(1)} & ... \\ 
...  & U^{(-2)} & U^{(-1)} & U^{(0)} & ...  \\ 
  \udots  & \vdots  & \vdots  & \vdots  & \ddots\\
\end{pmatrix},
\end{split}
\label{eq:Sambe_Keldysh_Density_functions}
\end{align}
where $\llangle x, m |\varrho | x, m^\prime \rrangle =  \rho^{(m^\prime - m)} (x)$   and $\llangle x, m | \mathcal{U} | x, m^\prime \rrangle =  U^{(m^\prime - m)} (x)$ with $\rho^{(n)}(x) = 1/T \int_0^T \, dt \, e^{i n \Omega t} \rho(x, t)$ and $U^{(n)}(x) = 1/T \int_0^T \, dt \, e^{i n \Omega t} U(x, t)$ being Fourier components of the time-periodic steady-state density and self-consistent Coulomb potential respectively.

Equation~(\ref{eq:Poison_Eq_FS}) is a time-independent Poisson equation but for matrix entries defined exploiting the Floquet-Sambe matrix structure. Thus, as anticipated, all calculations and algorithms developed for treating Coulomb interactions in stationary systems can  now be straightforwardly adapted for handling time-periodic drives. Note that in a practical calculation it is computationally more efficient to solve the Poisson equation for separate Fourier components of $\rho(x, t)$ and then represent the result in the form of~$\mathcal{U}(x)$. This is because $\varrho(x)$ contains a lot of repetitive information and it is necessary to stick to the Floquet-Sambe form only when performing matrix operations.

One always has to be aware that a complete electrodynamic theory has to include also a time-dependent magnetic field in addition to the Coulomb potential. In principle, this can be done by solving the time-periodic Schr\"{o}dinger equation self-consistently with Maxwell equations \cite{Wang},  which, as a matter of fact, can be represented in the Floquet-Sambe form in the same way as the Poisson equation, Eq.~(\ref{eq:Poison_Eq_FS}). However, the resulting magnetic field is usually very small and has negligible effect on the outcome. Thus we do not explicitly discuss this issue here.  

We finish this section with a brief discussion on the feasibility to adapt the approach for going beyond the mean-field treatment. As is well known, a Coulomb potential by itself does not account for effects coming from exchange interaction or correlations. These effects are usually incorporated in a single-particle time-independent picture by adapting concepts from density functional theory (DFT) \cite{Kohn1, Kohn2}.  Conventional DFT states that exchange and correlations can be included through an additional potential in the Hamiltonian, the so-called exchange-correlation (XC) functional $H_{\text{XC}}(x)$. This potential is a unique functional of the electron density $\rho(x)$ and can therefore be handled self-consistently. The exact form of this functional is unknown and it has to be guessed. Over the years quite a few approximations of the XC functional have been proposed and proved to be effective in predicting transport properties of various mesoscopic devices~\cite{Cohen_DFT, Kurth_DFT}, the simplest ones being the local density (LDA) and generalized gradient (GGA) approximations.

Interactions within a time-dependent formalism can be addressed with the help of time-dependent extension of the density functional theory (TDDFT) within which the density and XC functional now acquire explicit time-dependence~\cite{Kwok_TDDFT}. Even though Floquet variants of DFT have been proven to be not correct in general~\cite{Telnov_FDFT, Telnov_FDFT2, Chu _FDFT, Maitra _FDFT, Maitra _FDFT2}, Floquet theory can still be useful for some widely used approximations of the time-dependent XC functional. \text{Exploiting} Floquet %-Sambe design 
theory one can easily handle, for instance, adiabatic approximations popular within TDDFT, for example adiabatic local density (ALDA) and adiabatic generalized gradient (AGGA) approximations. These approximations use functionals from time-independent DFT and evaluate them at instantaneous density~\cite{Kwok_TDDFT}. This implies that for periodically driven systems such XC potentials can be brought to Floquet-Sambe form by considering separate Fourier components of the density. In this form time-dependent XC potentials will simply reduce to time-independent analogues but for objects within Floquet-Sambe space, implying that existing algorithms from time-independent DFT can be straightforwardly adapted.

\section{Self-consistent procedure}

\begin{figure} \centering
    \includegraphics[width=8.5cm,angle=0]{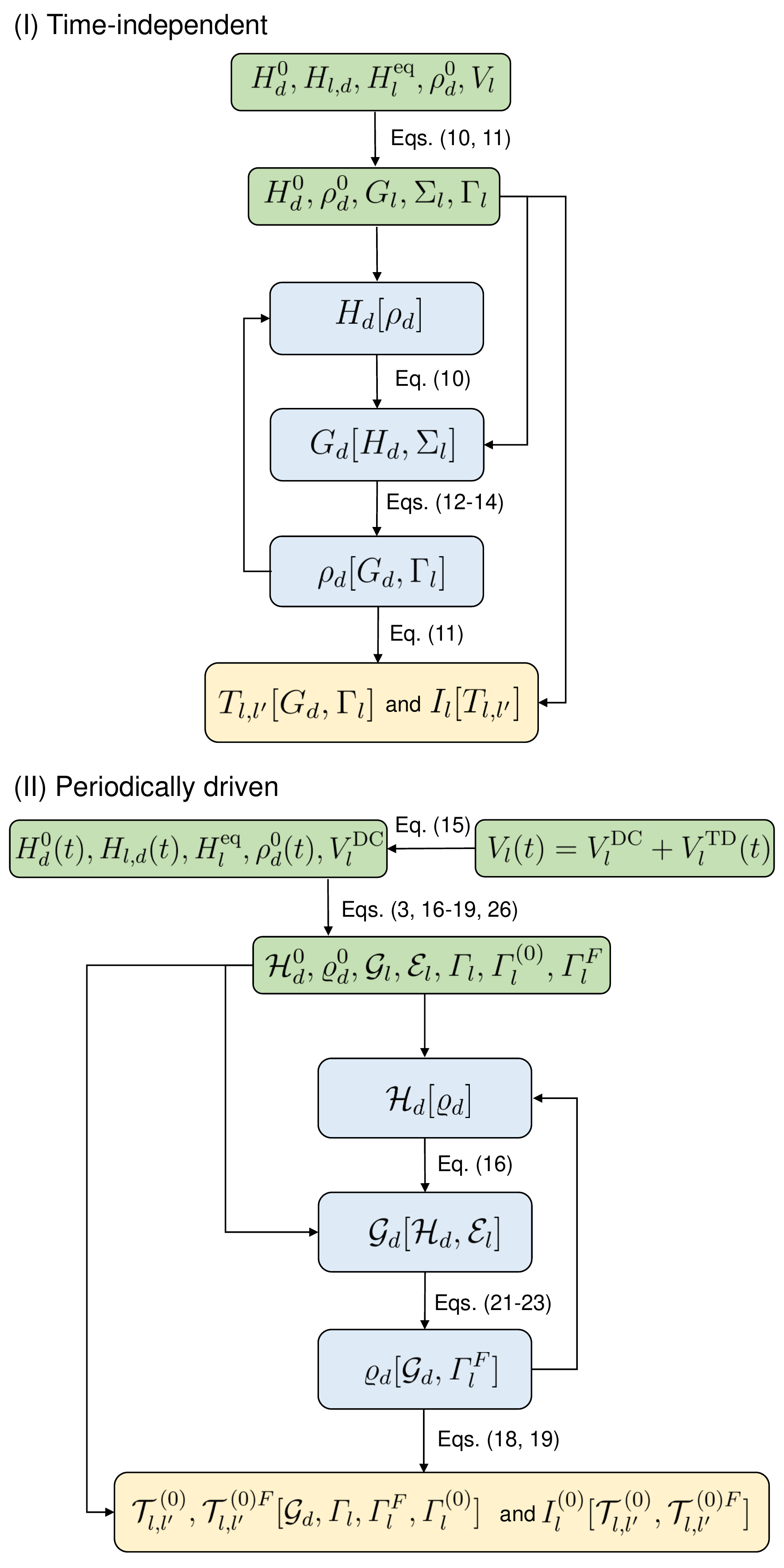}
    \caption{The recipes for calculating transport characteristics of (I) time-independent  and (II) periodically driven systems. In different colors we mark distinct stages of the algorithm: Initialization (green), self-consistent procedure (blue), and calculation of transport properties (yellow).}
     \label{fig2}
\end{figure}

In Fig. 2 we collect all the components discussed so far and present a unified computational scheme designed for describing transport in periodically driven systems.  For comparison, in the same figure we have also included a well-known recipe for performing time-independent transport calculations\cite{soft1, soft2}. The similarity between two schemes, the one for time-independent problems, the other for time-periodic ones, should be apparent. Nevertheless, there are a few modifications that one has to carry out for ``upgrading'' time-independent calculations to periodically driven ones. To a large extent, the modifications lie in a redefinition of matrices within frequency domain.

The computational scheme in Fig. 2 is divided into three stages: Initialization (green),  self-consistent procedure (blue), and calculation of transport properties (yellow). In the first stage one constructs matrices describing the central region and the leads as well as the auxiliary matrices used later in the procedure. When implemented in a computer code, this part is run just once and therefore it is expected to not require much computational time. The numerically most difficult part is the self-consistent procedure. Nevertheless, once all objects are constructed, the time-periodic and time-independent algorithms become essentially the same and therefore can be treated on the same footing. The matrices in the Floquet case are, however, typically larger and with their actual size depending on the cutoff used for truncating Floquet-Sambe space. Luckily, it is usually sufficient to use just a few Floquet blocks for reaching good accuracy of the result with reasonable computational time.  Lastly, the calculation of time-periodic transmission and current basically reduces to the well-known Landauer-B\"{u}ttiker expression within conventional time-independent Green's function formalism and by this it is relatively easy to implement.

To conclude this section, let us emphasize a number of aspects that are important to remember when executing the procedure in Fig.~2. They can be roughly divided into two categories: Conceptual and Computational. In the Conceptual category we include theoretical limitations of the method. Starting from the initialization stage we must remember that (i) the applied voltages are assumed to affect only phases of the states and to have no effect on their population. This is a widely used assumption but is expected to be correct only if the leads are good metals (which is often the case in experiments). (ii) The driven frequency and magnitude considered in the transport problem have to be comparable to the energy level separations of the undriven uncoupled system. Otherwise, the calculation will reduce to two well known limits, low and high frequency limits, and the introduced approach will not be relevant as one would be able to reduce the problem to the numerically much easier time-independent scenario. Moreover, in the low-frequency regime the Floquet-Sambe construction may be unpractical because truncation of the space may require an unreachable number of Floquet blocks or not be valid at all. The high-frequency limit may also be problematic because in transport calculations we often consider just part of the Hilbert space and high-frequency drives may take the system outside of that subspace. For example, in molecular electronics we often take into account just a few out-most bands and neglect core electrons and highly energetic levels.  Lastly, (iii) the Floquet bound states much below (above) the transport window are assumed to be filled (empty). While such populations may not always be formally correct for describing time-periodic steady states, they are however expected to be reasonably accurate for not too strong drives. If the distribution is different, or there are bound states close to the transport window, a different approach must be taken.

Now, we list some of the computational aspects of the algorithm that are important to keep in mind for efficient implementation. First of all, (i) in the initialization stage one would need to calculate the Floquet-Sambe surface Green's functions of the leads $\mathcal{G}_l(\omega)$ at complex-valued points used for calculating the equilibrium part of the density matrix $\mathcal{D}_d^\text{I}$, see Eq.~(\ref{eq:Density_FS_eq_trick}). The complex contour~$\mathcal{C}$, usually used for evaluating such integrals, can be found for example in~Refs.~$\text{[1, 2]}$. This implies that according to Eq.~(\ref{eq:Sambe_Keldysh_Green_functions_lead}) one would actually need conventional surface Green's functions of the leads, $G_l(\omega)$, evaluated at points $\omega \pm m\hbar\Omega \in \mathcal{C}$ with the integer $m$ bounded by the truncation limits of the Floquet-Sambe space. (ii)~There is a simple consistency check one may perform for verifying the computation: By setting all time-periodic amplitudes to zero, each Floquet-Sambe diagonal block (corresponding to some Floquet-Sambe indices $m = m^\prime$) has to exactly reproduce the time-independent result, and by this all blocks must be equal. This has to be true at every stage of the self-consistent procedure and therefore it is easily traceable. (iii) The truncation of the Floquet-Sambe space spoils the periodicity of the density $\varrho_d$ and it has to be always restored by hand after each iteration step. It is also more efficient to solve the Poisson equation for separate Fourier components of the density rather than for the full matrix $\varrho_d$.  (iv) The matrices within Floquet-Sambe formalism are in general much larger than in time-independent theory, even after truncation, and it may be helpful to optimize the matrix operations.

\section{Summary}

In this article we have conducted a comprehensive study on steady-state transport across periodically driven systems, addressed within a Green's function formalism combined with Floquet-Sambe theory. We have shown that all expressions for currents and densities essentially resemble the corresponding time-independent analogues widely used in numeric codes and analytic calculations. This important feature suggests that the proposed approach should be very useful and allow for direct transfer of computational algorithms and analytics developed for time-independent transport theory. In short, one just needs to redefine the operators within the Floquet-Sambe formalism, taking care of some straightforward complications coming from the periodic time-dependence. We have also addressed the numerical aspects of the proposed computational scheme and suggested simplified expressions for the currents and densities that are expected to be less costly to implement.

\section{Acknowledgements}

We are grateful to Henrik Johannesson and Oleksii Shevtsov for very useful and relevant discussions that significantly improved the article. This work was supported by the Swedish Research
Council through Grant No. 621-2014-5972.

\appendix

\section{Time-independent currents and density matrices}

In this appendix we collect technical details on the derivation of the expression for the dc current, Eq.~(\ref{eq:current_total_TI}), and density matrix, Eq.~(\ref{eq:Density_FS}). The basic idea is simple: one first calculates scattering states and then uses them for constructing the needed observables. Let us consider a bare mode $|l, E_k \rangle$ originating in lead~$l$ (eigenstate of $H_l$ with energy $E_k$ and quantum numbers $k$) and express the corresponding exact eigenmode of the full Hamiltonian $|u_{E_k} \rangle = |l, E_k \rangle + |\delta u_{E_k} \rangle$ using the Green's function~$G(E)$, defined in Eq.~(\ref{eq:Greeen_Def}). By direct substitution into the Schr\"{o}dinger equation we find that in the device region the mode is given by the relation $|u_{d,E_k} \rangle = G_d (E_k) H_{d,l}|l,E_k \rangle$, equivalent to the Lippmann-Schwinger equation. It follows that across lead~$l^\prime$ state $|u_{E_k} \rangle$ carries dc current 
\begin{align}
\begin{split}
i^k_{l \rightarrow l^\prime} = &  i\frac{e}{\hbar} [\langle u_{l^\prime, E_k} |H_{l^\prime,d}| u_{d, E_k}\rangle - \text{h.c.} ] \\ = & -\frac{e}{\hbar} \langle l, E_k|H_{l,d} G^\dagger_d \Gamma_{l^\prime} G_d H_{d,l}|l, E_k\rangle,
\end{split}
\label{eq:current single mode}
\end{align}
with $\Gamma_{l^\prime} = i(\Sigma_{l^\prime} - \Sigma_{l^\prime}^\dagger)$, $\Sigma_{l^\prime} = H_{d,{l^\prime}} \, G_{l^\prime} \, H_{{l^\prime}, d}$, and negative electron charge~$e$. The current is defined to be positive in the direction pointing from lead $l^\prime$ to the device region. 

According to the Landauer-B\"{u}ttiker formalism we assume that the $|u_{E_k} \rangle$ states are in equilibrium with a reservoir, implying the following expression for the current across~$l^\prime$ carried by modes originating in $l$:
\begin{align}
\begin{split}
I^\text{dc}_{l \rightarrow l^\prime} &=  \int_{-\infty}^{\infty} \, d\omega \, f_l(\omega - eV_l) \sum_k \delta(\omega-E_k) i^k_{l \rightarrow l^\prime}\\
                 &=  -\frac{e}{h} \int_{-\infty}^{\infty} \, d\omega \, f_l(\omega -eV_l) \text{Tr}  \left[ G^\dagger_d \Gamma_{l^\prime} G_d \Gamma_l \right](\omega), \\ 
\end{split}
\label{eq:current_LR_TI_Appendix}
\end{align}
where $ f_l(\omega - eV_l)$ is a Fermi-Dirac distribution function with chemical potential $\mu_l$ of reservoir $l$. To get rid of the delta function we have used the definition of a spectral function \text{$A_l = i[G_l - G_l^\dagger] = 2\pi \sum_k \delta(\omega -E_k) |l, E_k \rangle \langle l, E_k|$}. 

By exploiting conservation of current and summing up all current contributions one arrives at an expression for the total dc current flowing at terminal~$l$,
\begin{align}
\begin{split}
I^\text{dc}_{l} =  \frac{e}{h} \sum_{l^\prime \neq l} \int_{-\infty}^{\infty} \, d & \omega  \, \bigg(  \, T_{l, l^\prime} (\omega) \,  f_l(\omega - eV_l) \\
&- T_{l^\prime, l} (\omega)f_{l^\prime}(\omega - eV_{l^\prime}) \bigg),
\end{split}
\label{eq:current_total_TI_Appendix}
\end{align}
with $T_{l, l^\prime} = \text{Tr}  \left[ G^\dagger_d \Gamma_{l^\prime} G_d \Gamma_l \right]$. This expression can be simplified by noticing that $\sum_{l^\prime} \Gamma_{l^\prime}  = i\big((G^\dagger_d)^{-1} - G_d^{-1} + 2i\eta I_d\big)$ with infinitesimal $\eta >0$. It follows that $\sum_{l^\prime} \, T_{l, l^\prime} = \sum_{l^\prime} \, T_{l^\prime, l}$ and therefore~\cite{Datta}
\begin{align}
\begin{split}
I^\text{dc}_{l} =  \frac{e}{h} \sum_{l^\prime \neq l} \int_{-\infty}^{\infty} \, d  \omega  \, T_{l^\prime, l} (\omega) \,  \big( f_l(\omega - eV_l) - f_{l^\prime}(\omega - eV_{l^\prime}) \big).
\end{split}
\label{eq:current_total_TI_Appendix_2}
\end{align}

The density matrix in the device region, $D_d$, can be found in a similar way,
\begin{align}
\begin{split}
D_d &=  \sum_{l} \int_{-\infty}^{\infty} \, d\omega \, f_l(\omega - eV_l) \sum_k \delta(\omega-E_k)
| u_{d, E_k} \rangle \langle u_{d, E_k} |\\
& = \frac{1}{2\pi}\sum_{l} \int_{-\infty}^{\infty} \, d\omega \, f_l(\omega- eV_l) [G_d \Gamma_l G^\dagger_d](\omega).
\end{split}
\label{eq:density_TI_Appendix}
\end{align}

\section{Time-periodic currents and density matrices}

Here we generalize the approach outlined in \text{Appendix~A} to periodically driven systems and derive the corresponding expressions for currents and density matrices within Floquet-Sambe theory. All Fourier components of the time-periodic currents are considered here, not just the dc (rectified) component discussed in the main text.

Let us take a bare eigenstate $|l, E_k \rrangle$ of the time-independent lead $l$ (eigenstate of $\mathcal{H}_l$ with eigenvalue $E_k$ and set of quantum numbers~$k$) and use it to find the device region's part of the corresponding exact eigenmode $|u_{d, E_k} \rrangle = \mathcal{G}_d (E_k) \mathcal{H}_{d, l} | l, E_k \rrangle$. The steady-state current flowing across periodically driven systems is in general also time-periodic. Thus, the $n$-th Fourier component of the current carried by $|u_{E_k}\rrangle$ to lead $l^\prime$ through the coupling $H_{l^\prime, d}(t)$ is given by
\begin{align}
\begin{split}
&i^{k,n}_{l \rightarrow l^\prime} = i\frac{e}{\hbar}  \int_{0}^{T} \, \frac{dt}{T} \,e^{in \Omega t}  [\langle u_{l^\prime, E_k} (t) |H_{l^\prime, d}(t)| u_{d, E_k} (t)\rangle - \text{h.c.} ] \\
 =  & \ i\frac{e}{\hbar}  [\llangle u_{l^\prime, E_k} | \mathcal{K}_{n}\mathcal{H}_{ l^\prime, d}| u_{d, E_k}\rrangle - \llangle u_{d, E_k} |\mathcal{H}_{d, l^\prime} \mathcal{K}_{n}| u_{l^\prime, E_k}\rrangle ] \\ = & -i\frac{e}{\hbar} \llangle l, E_k| \mathcal{H}_{l,d}  \mathcal{G}^\dagger_d\mathcal{H}_{d, l^\prime} [\mathcal{K}_n \mathcal{G}_{l^\prime} - \mathcal{G}^\dagger_{l^\prime} \mathcal{K}_n] \mathcal{H}_{l^\prime, d}\mathcal{G}_d \mathcal{H}_{d,l}|l, E_k\rrangle\\
 =  & -\frac{e}{\hbar} \llangle l, E_k| \mathcal{H}_{l,d}  \mathcal{G}^\dagger_d \varGamma_{l^\prime (n)} \mathcal{G}_d \mathcal{H}_{d,l}|l, E_k\rrangle,
\end{split}
\label{eq:current single mode_Fq}
\end{align}
where $\varGamma_{l^\prime (n)} = i\mathcal{H}_{d, l^\prime} [\mathcal{K}_n \mathcal{G}_{l^\prime} - \mathcal{G}^\dagger_{l^\prime} \mathcal{K}_n] \mathcal{H}_{l^\prime, d}$. Here we have introduced a new operator ${\cal K}_n$ with matrix elements $\llangle i, m |\mathcal{K}_{n}| j, m^\prime \rrangle  = \int_{0}^{T} \, \frac{dt}{T} \langle i| e^{i(n + m^\prime - m) \Omega t} |j \rangle = \delta_{i,j}\delta_{m - m^\prime, \, n}$, which defines the identity matrix within Floquet-Sambe space shifted by~$n$ in the Floquet row index $m$. For $n = 0$ it is simply an identity matrix. As before the positive direction of the current ($i^{k,n}_{l \rightarrow l^\prime}>0$) is defined to point from lead $l^\prime$ to the device region. 

The leads are kept time-independent and therefore, according to Landauer-B\"{u}tiker formalism, the modes originating in them are in thermal equilibrium with the reservoirs (at chemical potentials $\mu_l$). As a result, the $n$-th Fourier component of the total current $I_{l \rightarrow l^\prime}(t)$ carried by modes from lead $l$ and passing across lead $l^\prime$ to the device region reads as
\begin{align}
\begin{split}
I^{(n)}_{l \rightarrow l^\prime} &= \int_{-\infty}^{\infty} d\omega f_l(\omega - eV_l) \sum_k \delta(\omega-E_k) i^{k,n}_{l \rightarrow l^\prime}\\
                 &= -\frac{e}{h} \int_{-\infty}^{\infty} d\omega f_l(\omega  - eV_l) \text{Tr}  \left[ \mathcal{G}^\dagger_d \varGamma_{l^\prime (n)} \mathcal{G}^{\phantom\dagger}_d \varGamma^{(0)}_{l} \right] (\omega).\\ 
\end{split}
\label{eq:current_LR_FS_Appendix}
\end{align}
Here $\varGamma^{(0)}_l = \mathcal{H}_{d, l}  [\mathcal{K}^0 \mathcal{A}_l] \mathcal{H}_{l, d}$ where $\llangle i, m |\mathcal{K}^0| j, m^\prime\rrangle  = \delta_{i,j} \delta_{m, 0} \delta_{m^\prime, 0}$ %is product of  $\mathcal{K}_n$ with $\delta_{m^\prime,0}$ 
and $\mathcal{A}_l = i(\mathcal{G}_l -  \mathcal{G}_l^\dagger)$ is the Floquet-Sambe spectral function of the $l$-th lead, cf.~Eq.~(\ref{eq:Spectral_Def_FS}). The delta functions $\delta_{m, 0}$ and $\delta_{m^\prime,0}$ appear due to the state overcounting in the definition of the Floquet-Sambe spectral function~$\mathcal{A}_l$: We have to sum over only distinct physical states, however, $\mathcal{A}_l$ contains all eigenmodes including redundant ones, cf.~Sec.~II.

The $n$-th Fourier component of the total current $I^{(n)}_{l}$ flowing across lead $l$ is found by adding up all contributions.
We need to be exceptionally careful here because the time-periodic current is in general sensitive to the position where it is calculated due to the displacement currents in the system. Thus, the current $I^{(n)}_{l}$ through the link $H_{l, d}(t)$ is given by
\begin{align}
\begin{split}
I_l^{(n)}  =  \sum_{l^\prime \neq l} \bigg(  &I^{(n)}_{l^\prime \rightarrow l} - I^{(n)}_{l \rightarrow l^\prime} \bigg) + \bigg(\frac{dQ^d_l}{dt}\bigg)^{(n)}\\ 
=\ \frac{e}{h} \sum_{l^\prime \neq l} & \int_{-\infty}^{\infty} \, d\omega  \,  \bigg( \,  
\mathcal{T}^{\, (n)}_{l, l^\prime} (\omega) f_{l}(\omega - eV_{l}) \\
-  &\mathcal{T}^{\, (n)}_{l^\prime, l} (\omega) f_{l^\prime}(\omega - eV_{l^\prime})  \bigg) +  \bigg(\frac{dQ^d_l(t)}{dt} \bigg)^{(n)},  \\ 
\end{split}
\label{eq:current_total_FS_appendix}
\end{align}
where $\mathcal{T}^{\, (n)}_{l, l^\prime} (\omega) = \text{Tr} \left[\mathcal{G}^\dagger_d \varGamma_{l^\prime  (n)}  \mathcal{G}^{\phantom\dagger}_d  \varGamma^{(0)}_{l} \right]$ and $\big( dQ^d_l(t)/dt\big)^{(n)}$ is the ${n}$-th Fourier component of time derivative of the total charge in the device region carried by the modes originating in lead~$l$, i.e. states $|u_{d, E_k} \rrangle$. The charge term is discussed in the end of the appendix. Here we have neglected by a contribution coming from the Floquet bound states since it can be made negligible by making the device region larger. Note that localized states do not carry any dc current and therefore the dc component is not affected by the bound states even if the device region is relatively small. Explicitly, the dc (rectified) current component reads as 
\begin{align}
\begin{split}
I_l^{(0)} =\ \frac{e}{h} \sum_{l^\prime \neq l}  \int_{-\infty}^{\infty} \, & d\omega  \,  \bigg( \,  
\mathcal{T}^{\, (0)}_{l, l^\prime} (\omega) f_{l}(\omega - eV_{l}) \\
-  &\mathcal{T}^{\, (0)}_{l^\prime, l} (\omega) f_{l^\prime}(\omega - eV_{l^\prime})  \bigg), \\ 
\end{split}
\label{eq:current_total_FS_appendix_dc}
\end{align}
where $\mathcal{T}^{\, (0)}_{l, l^\prime} = \text{Tr} \left[\mathcal{G}^\dagger_d \varGamma_{l^\prime}  \mathcal{G}^{\phantom\dagger}_d  \varGamma^{(0)}_{l} \right]$ represents the total transmission from lead $l$ to lead $l^\prime$. Here \text{$\varGamma_{l^\prime} = i(\mathcal{E}_{l^\prime} - \mathcal{E}_{l^\prime}^\dagger)$} with  $\mathcal{E}_{l^\prime} = \mathcal{H}_{d, l^\prime} \mathcal{G}_{l^\prime} \mathcal{H}_{l^\prime, d}$.

The time-periodic density matrix of the device region \text{$D_d(t)$} can be found in a similar way. To make the calculation more transparent we here split it into two steps. In the first step we simply rewrite $D_d(t)$ in Floquet-Sambe form, $\mathcal{D}_d$, with matrix elements
\begin{align}
\begin{split}
&\llangle i, m |\mathcal{D}_d | j, m^\prime \rrangle =  \llangle i, m |\sum_l\mathcal{D}^l_d | j, m^\prime \rrangle = \langle i| \sum_l D^{l \, (m^\prime - m)}_d |j\rangle\\
&= \sum_{l} \int d\omega f_l(\omega - eV_l) \, \sum_{k} \delta(\omega-E_k) \sum_n
\langle i| u^{n - m}_{d, E_k} \rangle \langle u^{n-m^\prime}_{d, E_k} |j \rangle,\\
\end{split}
\label{eq:density_TI_Floquet_Appendix}
\end{align}
where $D^{l}_d (t)$ is the time-periodic density matrix associated with the states originating in lead $l$. Here $D^{l \,(n)}_d$ and $|u^{n}_{d, E_k} \rangle$ denote Fourier components of the corresponding objects. At this stage it is important to remind ourselves that the time-periodic  modes~$|u(t)\rangle$ corresponding to physical states $|\psi(t)\rangle$ are not uniquely defined (cf. Sec.~II) and that the sum $\sum_k$ is only over physical states. In the second step, we make use of the redundancy in the representation of the physical states $|\psi(t)\rangle$, in particular of the property $|\psi_k(t)\rangle =  e^{-iE_kt/\hbar}|u_{E_k}(t)\rangle = e^{-iE_kt/\hbar - in\Omega t}|u_{E_k+n\hbar\Omega}(t)\rangle$. This property allows us to relate Floquet components of time-periodic modes associated with the same physical state as  $|u^{n + m}_{E_k} \rangle =|u^{m}_{E_k + n\hbar\Omega} \rangle$. From Eq.~(\ref{eq:density_TI_Floquet_Appendix}) we then obtain 
\begin{align}
\begin{split}
&\ \ \ \, \llangle i, m |\mathcal{D}_d | j, m^\prime \rrangle \\
&= \sum_{l} \int d\omega f_l \, \sum_{k,n} \delta(\omega-E_k) 
\langle i| u^{-m}_{d, E^{n}_k} \rangle \langle u^{-m^\prime}_{d, E^n_k} |j \rangle\\
&= \llangle i, m| \left[\sum_{l,k,n} \int d\omega f_l \, \delta(\omega-E_{k}) 
 |u_{d, E^n_k} \rrangle \llangle u_{d, E^n_k}| \right]|j, m^\prime \rrangle\\
&= \llangle i, m| \left[\sum_{l,n} \int d\omega f_l  \mathcal{G}_d (\omega_n) \mathcal{H}_{d,l} \mathcal{A}^{n}_l(\omega) \mathcal{H}_{l,d} \mathcal{G}^\dagger_d (\omega_n) \right]|j, m^\prime \rrangle\\
& = \frac{1}{2\pi} \llangle i, m| \left[ \sum_{l} \int_{-\infty}^{\infty} \, d\omega \, [\mathcal{G}_d \varGamma^F_l \mathcal{G}^\dagger_d](\omega) \right]|j, m^\prime \rrangle.\\
\end{split}
\label{eq:density_TI_Floquet_Appendix2}
\end{align}
Here $f_l$ is a short notation for $f_l(\omega -eV_l)$, the state $|u_{d, E^n_{k}} \rrangle$ denotes $|u_{d, E_{k}} \rrangle$ shifted by $n$ columns in Floquet index, i.e  $\llangle i, m|u_{d, E^n_{k}} \rrangle = \llangle i, m - n|u_{d, E_{k}} \rrangle$ with $E^n_k = E_k + n\hbar\Omega$, $\omega_n$ is short for $\omega + n \hbar \Omega$, and the matrix elements $\llangle i, m|\mathcal{A}^{n}_l(\omega)|j, m^\prime \rrangle = \langle j|A_l(\omega)|i \rangle \delta_{n, m} \delta_{n, m^\prime}$ represent the $l$-th lead spectral function placed into the Floquet-Sambe zero matrix exactly on the $n$-th diagonal slot. The final expression contains the operator $\varGamma^F_l (\omega) = \mathcal{H}_{d,l} [\mathcal{A}_l \mathcal{F}_l] \mathcal{H}_{l,d}$ with $\llangle i, m|\mathcal{F}_l(\omega)|j, m^\prime \rrangle = f_l (\omega - eV_l - m \hbar\Omega ) \delta_{m, m^\prime} \, \delta_{i, j}$ \text{being} a representation of the Fermi-Dirac distribution function \text{$f_l (\omega - eV_l)$} within Floquet-Sambe space \cite{Cao1, Eissing}. 

The total charge $Q^d_l(t)$ in the device region composed of occupied states coming from lead $l$ is found by computing the trace of the density matrix $Q^d_l(t) = e \, \text{Tr} [D^l_d(t)]$. It follows that~$\mathcal{Q}^d_l$, the Floquet-Sambe analogue to $Q^d_l(t)$, is given~by  
\begin{align}
\begin{split}
&\llangle m |\mathcal{Q}^d_l|  m^\prime \rrangle =  \frac{e}{2\pi} \displaystyle\text{Tr} \, \llangle i, m| \left[ \int_{-\infty}^{\infty} \, d\omega \, [\mathcal{G}_d \varGamma^F_l \mathcal{G}^\dagger_d](\omega) \right]|j, m^\prime \rrangle,
\end{split}
\label{eq:charge_Floquet-Sambe_appendix}
\end{align}
where the trace is taken over $i$ and $j$ indices. The derivative $dQ^d_l(t)/dt$ is obtained through $\llangle m |dQ^d_l(t)/dt|  m^\prime \rrangle = \llangle m |-i\Omega (m^\prime - m)\mathcal{Q}^d_l|  m^\prime \rrangle = -i\Omega \llangle m |(\mathcal{Q}^d_l\mathcal{M} - \mathcal{M}\mathcal{Q}^d_l)|  m^\prime \rrangle$ with $\llangle m| \mathcal{M}|  m^\prime \rrangle = m \, \delta_{m,m^\prime}$. The $n$-th Fourier component of $dQ^d_l(t)/dt$, used in Eq.~(\ref{eq:current_total_FS_appendix}), is then simply given by the $(m, \, m^\prime)$ element of the matrix $-i\Omega[\mathcal{Q}^d_l\mathcal{M} - \mathcal{M}\mathcal{Q}^d_l]$ with $m^\prime - m = n$. 

\section{The simplified expression for the dc component of the time-periodic current}

The unbounded integration interval in Eq.~(\ref{eq:current_total_FS_appendix_dc}) significantly complicates the computations: The calculation of the dc current is still expected to be manageable, however, requires much more computational resources in comparison to the time-independent case. In what follows we transform this equation into a more computationally friendly form, cf. Eq.~(\ref{eq:current_total_FS_TRS_final_short}). To proceed we first decompose 
the expression for the total dc transmission $\mathcal{T}^{\, (0)}_{l, l^\prime} = \text{Tr} \left[ \mathcal{G}^{\phantom\dagger}_d \varGamma^{(0)}_{l} \mathcal{G}^\dagger_d \varGamma_{l^\prime} \right]$ into Floquet-Sambe blocks. It is done in the following way: 
\begin{align}
\begin{split}
\mathcal{T}^{\, (0)}_{l, l^\prime} =  \sum_{\textbf{m}} \mathcal{T}^{\, (0) \, \textbf{m}}_{l, l^\prime \, \text{I}}, 
\end{split}
\label{eq:Transmision_TRS_FS_Components}
\end{align}
with 
\begin{align}
\begin{split}
& \mathcal{T}^{\, (0) \, \textbf{m}}_{l, l^\prime \, \text{I}} = \text{Tr}  \bigg [ [\mathcal{G}_d (\omega)]^{(m_1, m_2)} \mathcal{H}^{(m_2, 0)}_{d, l} [\mathcal{A}_{l} (\omega)]^{(0, 0)} \mathcal{H}^{(0, m_3)}_{l, d} \\
& [\mathcal{G}^\dagger_d (\omega)]^{(m_3, m_4)}  \mathcal{H}^{(m_4, m_5)}_{d, l^\prime} [\mathcal{A}_{l^\prime}(\omega)]^{(m_5, m_5)} \mathcal{H}^{(m_5, m_1)}_{l^\prime, d} \bigg ].  \\
\end{split}
\label{eq:Transmision_TRS_FS_Components_2}
\end{align}
Here the trace is taken over conventional basis states~$|i \rangle$, the integer upper indices $m_\alpha$ with $\alpha = 1, ..., 5$ label the corresponding Floquet-Sambe blocks, and $\textbf{m} = (m_1, m_2, m_3, m_4, m_5)$. The sum in Eq.~(\ref{eq:Transmision_TRS_FS_Components}) is over all integers $m_\alpha$, i.e. over all configurations $\textbf{m}$. Let us also define the following quantity: 
\begin{align}
\begin{split}
& \mathcal{T}^{\, (0) \, \textbf{m}}_{l, l^\prime \, \text{II}} = \text{Tr}  \bigg [ [\mathcal{G}^\dagger_d (\omega)]^{(m_1, m_2)} \mathcal{H}^{(m_2, 0)}_{d, l} [\mathcal{A}_{l} (\omega)]^{(0, 0)} \mathcal{H}^{(0, m_3)}_{l, d} \\
& [\mathcal{G}_d (\omega)]^{(m_3, m_4)}  \mathcal{H}^{(m_4, m_5)}_{d, l^\prime} [\mathcal{A}_{l^\prime}(\omega)]^{(m_5, m_5)} \mathcal{H}^{(m_5, m_1)}_{l^\prime, d} \bigg ].  \\
\end{split}
\label{eq:Transmision_TRS_FS_Components_3}
\end{align}
We now use the periodicity of the Floqet-Sambe matrices, cf. Sec.~II, and shift down every term in Eq.~(\ref{eq:Transmision_TRS_FS_Components_3}) by $m_5$ Floquet-Sambe rows and columns. By the Floquet-Sambe periodic property this shift will not change the matrices except by an additional shift \text{$\omega \rightarrow \omega - m_5 \hbar \Omega$} in all the quantities that are \text{$\omega$-dependent}. Therefore, by exploiting the cyclic property of the trace we get
\begin{align}
\begin{split}
\mathcal{T}^{\, (0) \, \textbf{m}}_{l, l^\prime \, \text{II}} (\omega) = \mathcal{T}^{\, (0) \, \overline{\textbf{m}}}_{l^\prime, l \, \text{I}} (\omega - m_5 \hbar \Omega), 
\end{split}
\label{eq:Transmision_TRS_FS_Components_Eq}
\end{align}
where the upper indices are $\textbf{m} = (m_1, m_2, m_3, m_4, m_5)$ and $\overline{\textbf{m}} = (m_3 - m_5, m_4 - m_5, m_1 -m_5, m_2 - m_5, -m_5)$. Note that all possible configurations of $\overline{\textbf{m}}$ repeat all possible configurations of $\textbf{m}$. By combining Eqs.~(\ref{eq:current_total_FS_appendix_dc}, \ref{eq:Transmision_TRS_FS_Components}-\ref{eq:Transmision_TRS_FS_Components_3}) we arrive at the following expression for the dc current:
\\
\\
\begin{align}
\begin{split}
I_l^{(0)} =\ \frac{e}{h} \sum_{l^\prime, \, \textbf{m}}  \int_{-\infty}^{\infty} \, & d\omega  \,  \bigg( \, 
\mathcal{T}^{\, (0) \, \textbf{m}}_{l, l^\prime \, \text{I}} (\omega) f_{l}(\omega - eV_{l}) \\
-  &\mathcal{T}^{\, (0) \, \overline{\textbf{m}}}_{l^\prime, l  \, \text{I}} (\omega) f_{l^\prime}(\omega - eV_{l^\prime})  \bigg) \\ 
=\ \frac{e}{h} \sum_{l^\prime , \, \textbf{m}}  \int_{-\infty}^{\infty} \, & d\omega  \,  \bigg( \, 
\mathcal{T}^{\, (0) \, \textbf{m}}_{l, l^\prime \, \text{I}} (\omega) f_{l}(\omega - eV_{l}) \\
-  &\mathcal{T}^{\, (0) \, \textbf{m}}_{l, l^\prime \, \text{II}} (\omega + m_5 \hbar \Omega) f_{l^\prime}(\omega - eV_{l^\prime})  \bigg) \\ 
=\ \frac{e}{h} \sum_{l^\prime, \, \textbf{m}}  \int_{-\infty}^{\infty} \, & d\omega  \,  \bigg( \, 
\mathcal{T}^{\, (0) \, \textbf{m}}_{l, l^\prime \, \text{I}} (\omega) f_{l}(\omega - eV_{l}) \\
-  &\mathcal{T}^{\, (0) \, \textbf{m}}_{l, l^\prime \, \text{II}} (\omega) f_{l^\prime}(\omega - eV_{l^\prime} - m_5 \hbar \Omega)  \bigg).
\end{split}
\label{eq:current_total_FS_appendix_TRS_part}
\end{align}

By making use of the summation over all possible configurations~$\textbf{m}$ and the definition of  $\varGamma^F_{l}$ introduced in Appendix~B, we retrieve the following Floquet-Sambe representation of the dc current $I_l^{(0)}$:
\begin{align}
\begin{split}
I_l^{(0)} =\ \frac{e}{h} \sum_{l^\prime}  \int_{-\infty}^{\infty} \,  d\omega  \,  \bigg( \, 
\mathcal{T}^{\, (0)}_{l, l^\prime} (\omega) f_{l}(\omega - eV_{l}) \, - \, &\mathcal{T}^{\, (0) \, F}_{l, l^\prime} (\omega)  \bigg),
\end{split}
\label{eq:current_total_FS_appendix_TRS_final_short}
\end{align}
where the transmissions are $\mathcal{T}^{\, (0)}_{l, l^\prime} = \text{Tr} \left[\mathcal{G}^\dagger_d \varGamma_{l^\prime}  \mathcal{G}^{\phantom\dagger}_d  \varGamma^{(0)}_{l} \right]$ and $\mathcal{T}^{\, (0) \, F}_{l, l^\prime} = \left[\mathcal{G}^{\phantom\dagger}_d \varGamma^F_{l^\prime}  \mathcal{G}^\dagger_d  \varGamma^{(0)}_{l} \right]$.

We notice that $\sum_{l^\prime} \, \mathcal{T}^{\, (0)}_{l, l^\prime} f_{l}(\omega - eV_{l}) \simeq \sum_{l^\prime} \, \mathcal{T}^{\, (0) \, F}_{l, l^\prime}$ in the energy region where $\varGamma^F_{l^\prime} \simeq \varGamma_{l^\prime}$ for all $l^\prime$, i.e. in the interval where the Fermi function within the truncated Floquet-Sambe space, $\mathcal{F}_{l^\prime}$, is close to identity for all~$l^\prime$. This equality directly follows from the definitions of $\mathcal{T}^{\, (0)}_{l, l^\prime}$ and $\mathcal{T}^{\, (0) \, F}_{l, l^\prime}$ combined with the identity $\sum_{l^\prime} \varGamma_{l^\prime}  = i\big((\mathcal{G}^\dagger_d)^{-1} - \mathcal{G}_d^{-1} +2i\eta I_d \big)$, where $\eta$ is infinitesimal and positive. Therefore, the integral in  Eq.~(\ref{eq:current_total_FS_appendix_TRS_final_short}) has to be evaluated only over the energy region where the Fermi-Dirac matrix $\mathcal{F}_l$ is neither close to identity nor to zero matrix.

\end{document}